\input harvmac
\noblackbox
\input epsf


\newcount\figno
\figno=0
\def\fig#1#2#3{
\par\begingroup\parindent=0pt\leftskip=1cm\rightskip=1cm\parindent=0pt
\baselineskip=11pt

\global\advance\figno by 1
\midinsert
\epsfxsize=#3
\centerline{\epsfbox{#2}}
\vskip 12pt
\centerline{\vbox{{\bf Figure \the\figno:} #1}}\par
\endinsert\endgroup\par}
\def\figlabel#1{\xdef#1{\the\figno}}
\overfullrule=0pt


\def\underarrow#1{\vbox{\ialign{##\crcr$\hfil\displaystyle
 {#1}\hfil$\crcr\noalign{\kern1pt\nointerlineskip}$\longrightarrow$\crcr}}}
%
\def\tilde{\widetilde}
\def\bar{\overline}

\def\np#1#2#3{Nucl. Phys. {\bf B#1} (#2) #3}
\def\pl#1#2#3{Phys. Lett. {\bf #1B} (#2) #3}

\def\physrev#1#2#3{Phys. Rev. {\bf D#1} (#2) #3}

\def\jhep#1#2#3{J. High Energy Phys. {\bf #1} (#2) #3}
%

\font\cmss=cmss10
\font\cmsss=cmss10 at 7pt
\def\rlx{\relax\leavevmode}
\def\inbar{\vrule height1.5ex width.4pt depth0pt}
\def\IC{\relax\,\hbox{$\inbar\kern-.3em{\rm C}$}}
\def\IN{\relax{\rm I\kern-.18em N}}
\def\IP{\relax{\rm I\kern-.18em P}}
\def\IR{\relax{\rm I\kern-.18em R}}
\def\IC{{\relax\hbox{$\inbar\kern-.3em{\rm C}$}}}
\def\IZ{\relax\ifmmode\mathchoice
{\hbox{\cmss Z\kern-.4em Z}}{\hbox{\cmss Z\kern-.4em Z}}
{\lower.9pt\hbox{\cmsss Z\kern-.4em Z}}
{\lower1.2pt\hbox{\cmsss Z\kern-.4em Z}}\else{\cmss Z\kern-.4em
Z}\fi}
\def\IH{\relax{\rm I\kern-.18em H}}
\def\ZZ{\rlx\leavevmode\ifmmode\mathchoice{\hbox{\cmss Z\kern-.4em Z}}
 {\hbox{\cmss Z\kern-.4em Z}}{\lower.9pt\hbox{\cmsss Z\kern-.36em Z}}
 {\lower1.2pt\hbox{\cmsss Z\kern-.36em Z}}\else{\cmss Z\kern-.4em
 Z}\fi}
\def\narrowplus{\kern -.04truein + \kern -.03truein}
\def\narrowminus{- \kern -.04truein}
\def\narrowminussub{\kern -.02truein - \kern -.01truein}

\def\ext{{\rm Ext}}
\def\hom{{\rm Hom}}

\def\frac#1#2{{#1\over #2}}

\def\CN{{\cal N}}

\def\BR{\IR}
\def\BZ{\IZ}

\def\Bone{{\bf 1}}

\def\IZ{\relax\ifmmode\mathchoice
{\hbox{\cmss Z\kern-.4em Z}}{\hbox{\cmss Z\kern-.4em Z}}
{\lower.9pt\hbox{\cmsss Z\kern-.4em Z}}
{\lower1.2pt\hbox{\cmsss Z\kern-.4em Z}}\else{\cmss Z\kern-.4em
Z}\fi}

\def\RP#1{{\BR {\rm P}^{#1}}}
\def\tZ{\tilde{\BZ}}

\def\tO{{\tilde{O2}^+}}

%
%
\def\eqnn#1{\xdef #1{(\secsym\the\meqno)}\writedef{#1\leftbracket#1}%
\global\advance\meqno by1\wrlabeL#1}
\def\eqna#1{\xdef #1##1{\hbox{$(\secsym\the\meqno##1)$}}
\writedef{#1\numbersign1\leftbracket#1{\numbersign1}}%
\global\advance\meqno by1\wrlabeL{#1$\{\}$}}
\def\eqn#1#2{\xdef #1{(\secsym\the\meqno)}\writedef{#1\leftbracket#1}%
\global\advance\meqno by1$$#2\eqno#1\eqlabeL#1$$}

\nref\rseibergsqcd{N.~Seiberg, ``Electric-Magnetic Duality in 
Supersymmetric
Nonabelian Gauge Theories,'' \np{435}{1995}{129}, hep-th/9411149.}

\nref\rmirror{N.~Seiberg and K.~Intriligator, ``Mirror Symmetry in 
Three-Dimensional
Gauge Theories,'' \pl{387}{1996}{513}, hep-th/9607207.}

\nref\rsethi{S.~Sethi, ``A Relation Between $N=8$ Gauge Theories in 
Three Dimensions,'' hep-th/9809162.}

\nref\porrati{M. Porrati and A. Zaffaroni, ``M Theory Origin of Mirror
Symmetry in Three-Dimensional Gauge Theories,'' \np{490}{1997}{107},
hep-th/9611201.}

\nref\rwitten{E.~Witten, ``Baryons and Branes in Anti de Sitter 
Space,'' \jhep{9807}{1998}{006}, hep-th/9805112.}

\nref\vafan{M.~Bershadsky, Z.~Kakushadze and C.~Vafa, ``String 
Expansion as Large N Expansion of Gauge Theories'', \np{523}{1998}{59},
hep-th/9803076.}

\nref\rseiberg{N.~Seiberg, ``Notes on Theories with Sixteen 
Supercharges,'' hep-th/9705117.}

\nref\pg{E.G.~Gimon and J.~Polchinski, ``Consistency Conditions for 
Orientifolds and D-Manifolds'', \physrev{54}{1996}{1667}, hep-th/9601038.}

\nref\joet{J.~Polchinski, ``Tensors from K3 Orientifolds'', 
\physrev{55}{1997}{6423}, hep-th/9606165.} 

\nref\blsspw{M.~Berkooz, R.G.~Leigh, J.~Polchinski, J.H.~Schwarz and 
E.~Witten, ``Anomalies, Dualities, and Topology of D=6 N=1 Superstring 
Vacua'', \np{475}{1996}{115}, hep-th/9605184.}

\nref\ken{K. Intriligator, ``RG Fixed Points in Six Dimensions via 
Branes at Orbifold Singularities'', \np{496}{1997}{177}, hep-th/9702038.} 

\nref\aspinwall{P.S.~Aspinwall, ``Enhanced Gauge Symmetries and K3
Surfaces,'' \pl{357}{1995}{329}, hep-th/9507012.}

\Title{\vbox{\hbox{hep-th/9810257}\hbox{IASSNS-HEP-98/89}}}
{\vbox{\centerline{New IR Dualities in Supersymmetric Gauge Theory}
\centerline{}
\centerline{in Three Dimensions}}}
\smallskip
\centerline{Micha Berkooz\footnote{$^1$} {berkooz@sns.ias.edu} and
Anton Kapustin\footnote{$^2$} {kapustin@sns.ias.edu} }
\vskip 0.12in
\medskip\centerline{\it School of Natural Sciences}
\centerline{\it Institute for Advanced Study}\centerline{\it
Princeton, NJ 08540, USA}

\vskip 1in

We present nontrivial examples of $d=3$ gauge theories with sixteen
and eight supercharges which are infrared dual at special points in the
moduli space.  This duality is distinct from mirror symmetry. To
demonstrate duality we construct the gauge theories of interest using
D2-branes and orientifolds and then consider their lift to
M-theory. We also discuss the strong coupling limit of orientifold
two-planes and orbifolds of orientifold six-planes.

\vfill\eject
\vskip 0.1in

\newsec{Introduction}

The term ``duality'' in field theory has several
meanings. Electric-magnetic duality is a variable transformation in
free abelian gauge theories exchanging the field strength $F$ and its
dual $*F$. Strong-weak coupling duality applies to theories which have
exactly marginal parameters (``couplings'').  It is an equivalence
between a strongly coupled theory and another theory which is weakly
coupled.  For example, an $\CN=4$ $d=4$ gauge theory with gauge group
$G$ and complexified gauge coupling $\tau$ is equivalent to an $\CN=4$
theory with gauge group $\hat{G}$ (the Langlands dual of $G$) and
coupling $-1/\tau$. Yet another type of duality is infrared (IR)
duality. Two theories are called IR dual if they flow to the same
infrared fixed point. Nontrivial examples in $d=4$ are some ${\cal
N}=1$ Seiberg dual pairs \rseibergsqcd.  Mirror theories in $d=3$
\rmirror\ provide another example of IR duality.

In this letter we present new examples of IR dual theories in $d=3$
with sixteen and eight supercharges. This duality is distinct from
mirror symmetry: in all our examples the Coulomb branch moduli of the
dual theories are identified, while mirror symmetry exchanges Coulomb
and Higgs branches. In the less interesting cases IR duality is
visible already at the classical level.  For example, at a generic
point in the moduli space an $\CN=8$ theory with gauge group $G$ flows
to a free $\CN=8$ theory with gauge group $U(1)^r$ where $r={\rm rank}
G$.  Thus any two $\CN=8$ theories with gauge groups of equal rank are
IR dual at a generic point in the moduli space. The dual pairs
discussed in this letter are nontrivial in the sense that their IR
equivalence cannot be seen classically.

In Section 2 we show that at a special point in the moduli space an
$\CN=8$ theory with gauge group $G=Sp(2N)$ is dual to an $\CN=8$
theory with $G=O(2N)$, while at another special point it is dual to an
$\CN=8$ theory with $G=O(2N+1)$. The latter duality has been
previously derived in
\rsethi. These dualities hold in the vicinity of orbifold singularities 
of the
moduli space, where classically the full gauge symmetry is
restored. Consequently, IR equivalence at these points is a
quantum-mechanical phenomenon.

In Section 3 we present an example of IR duality for $\CN=4$ theories
in $d=3$. We show that at a special point in the moduli space a
$U(2N)$ theory with two fundamentals and two antisymmetric tensors is
dual to an $Sp(2N)\times Sp(2N)$ theory with a hypermultiplet in the
$({\bf 2N, 2N})$, a hypermultiplet in the $({\bf 2N,1})$, and a
hypermultiplet in the $({\bf 1, 2N})$. This duality is not visible
classically.

To show infrared equivalence we construct the theories of interest
using D2-branes and orientifolds and then consider the limit of strong
IIA coupling. 
The main idea is that in $d=3$
theories the RG flow is influenced by the VEV of the dual photon. This
means that a IIA singularity probed by D2-branes can be resolved into
several singularities along the M-theory circle. Theories which look
very different in the UV can differ in the IR only by the number and type
of singular points. Since the critical IR behavior
is determined by the local M-theory geometry, different UV theories
may flow to the same IR theory at special points on the moduli space.
Similar methods
have been used to demonstrate mirror symmetry \porrati.

In the case of $\CN=8$ theories one needs to understand the strong
coupling limit of various O2 planes.  The lift of O2 planes to
M-theory is studied in Section 2; our discussion there overlaps with
that in \rsethi.  In the case of $\CN=4$ theories the relevant
M-theory background is an orbifold $\BR^4/\BZ_2\times \BR^4/\BZ_2$
which we study in subsection 3.2.

\newsec{$\CN=8$ supersymmetric gauge theories in three dimensions}

\subsec{IIA brane configurations and the classification of orientifold
two-planes}

Consider $N$ D2-branes parallel to an orientifold 2-plane. The
low-energy theory on the D2 worldvolume is an $\CN=8$ $d=3$ gauge
theory. The gauge group $G$ depends on the choice of the orientifold
projection. If the RR charge of the O2 plane is positive the gauge
group is $Sp(2N)$; if it is negative then $G=O(2N)$. We will call the
former an O2$^+$ plane, and the latter an O2$^-$ plane. Their RR
charges are $1/8$ and $-1/8$, respectively, in the units where the
charge of a D2-brane is $1$. In the case of an O2$^-$
plane one can also consider adding half of a D2-brane stuck at the
fixed point.  Following \rsethi\ we denote this orientifold plane by
$\tilde{O2}^+$; its RR charge is $3/8$. The low-energy theory on
D2-branes near an $\tilde{O2}^+$ plane has gauge group $G=O(2N+1)$.

The above classification of O2 planes was based on the RR
charge. We can also consider a topological classification similar to
that in \rwitten. One asks whether there are inequivalent ways to
quantize a p-brane propagating away from the orientifold
plane. Requiring that the p-brane be away from the fixed plane is
equivalent to replacing the manifold $\BR^7/\BZ_2$ with $\BR\times
\RP6$ which is homotopic to $\RP6$. Inequivalent ways of assigning phases 
to disconnected sectors in the p-brane path integral are classified by
appropriate (co)homology groups of $\RP6$. The p-branes of interest to
us are fundamental strings and D2-branes. Inequivalent ways of choosing
phases for a fundamental string path integral are classified by
$\hom(H_2(\RP6,\tZ),U(1))=\hom(\BZ_2,U(1))=\BZ_2$.\foot{Following
\rwitten\ we denote by $\tZ$ a locally constant sheaf of integers whose 
sections change sign when going around a noncontractible loop in
$\RP6$.}  Equivalently, the possible fluxes of the NS 3-form field
strength $H=dB$ are classified by $H^3(\RP6,\tZ)=\BZ_2$. Since this is
pure torsion, the nontrivial cohomology class can be realized by a
2-form $B$ with vanishing field strength $H$. The only effect of this
$B$-field is to multiply certain contributions to the stringy path
integral by $-1$. More precisely, turning on the $B$-field changes the
sign of the contribution of worldsheets wrapping the generator of
$H_2(\RP6,\tZ)$ (these have $\RP2$ topology). Recall now that string
perturbation theory for D-branes is equivalent to large $N$ expansion
in gauge theory \vafan. In the large $N$ limit the difference between
orthogonal and symplectic gauge groups is precisely the sign of the
$\RP2$ contributions. Thus we conclude that O2$^-$ and $\tO$ on one
side and O2$^+$ on the other side are distinguished by discrete
torsion in $H^3(\RP6,\tZ)$.  We will argue below that it is O2$^+$
which has nontrivial torsion.

The other type of p-brane we consider is D2-brane. The choice of
phases in the path integral for D2-branes is classified by
$\hom(H_3(\RP6,\BZ),U(1))=\hom(\BZ_2,U(1))=\BZ_2$. By universal
coefficient formulas $H^4(\RP6,\BZ)=\ext(H_3(\RP6,\BZ),\BZ)=\BZ_2$, so
one can also think of the phase ambiguity as the freedom to choose a
cohomology class of the RR 4-form field strength $G^{(4)}$. Since the
cohomology is pure torsion, it again can be represented by a $C^{(3)}$
with vanishing field strength. We believe that that O2$^-$ and $\tO$
have different $H^4(\RP6,\BZ)$ torsion, but we will not try to prove
it here.

\subsec{Infrared limit and the lift to M-theory}

In the extreme infrared the gauge theories in question are described
by $N$ M2-branes probing a $d=11$ supergravity background.  The moduli
space of $N$ D2-branes parallel to an orientifold two-plane is the
moduli space of an appropriate $\CN=8$ gauge theory. Standard
arguments
\rseiberg\ show that this moduli space is an orbifold 
$Sym((\BR^7\times S^1)/\BZ_2)^N$, hence all O2 planes lift to an
M-theory orbifold $(\BR^7 \times S^1)/\BZ_2$.  Let us denote the
coordinates on $\BR^7$ by $x_1,\ldots,x_7$ and the coordinate on $S^1$
by $\sigma$, with the convention that $\sigma$ has period $2\pi$.
There are two points which are left invariant by the orbifold action,
$x_1=\cdots=x_7=\sigma=0$, and $x_1=\cdots=x_7=0,\ \sigma=\pi$. We
denote them by $p_+$ and $p_-$. We will denote by $m_+$ (resp. $m_-$)
the point in the gauge theory moduli space which corresponds to all
M2-branes sitting at $p_+$ (resp. $p_-$). Other singular points in the
moduli space are obtained if we put $k>0$ of the membranes at $p_+$
and the rest at $p_-$. It is clear that for $k>0$ the resulting SCFT
decomposes into a product of two independent SCFT's, one at $p_+$ and
another at $p_-$. Both of these SCFT's can be also obtained by
starting from a gauge group of lower rank and putting all membranes at
the same point. Thus without loss of generality we may concentrate on
the situation when all membranes are at the same point.

The neighborhood of either $p_+$ or $p_-$ looks like
$\BR^8/\BZ_2$. There are in fact two different $\BR^8/\BZ_2$ orbifolds
of M-theory differing by their membrane charge and the cohomology
class of the 4-form field strength $G$. To see this, one can study the
possibility of introducing phases in the path integral for
membranes. As before, we delete the singular point of $\BR^8/\BZ_2$
which makes it homotopic to $\RP7$. The phases are classified by
$\hom(H_3(\RP7,\BZ),U(1))=\BZ_2$, which by universal coefficient
formulas is related to $H^4(\RP7,\BZ)=\BZ_2$.  Nontrivial
$H^4(\RP7,\BZ)$ torsion means that in the membrane path integral the
contribution of membranes wrapped on the cycles homologous to $\RP3$
linearly embedded in $\RP7$ has an extra $-1$. We will call the
$\BR^8/\BZ_2$ orbifolds with and without torsion B and A-orbifolds,
respectively.

It was shown in \rsethi\ that the membrane charge of the A-orbifold
(the one without torsion) is $-1/16$, while that of the B-orbifold is
$3/16$. Using this knowledge it is easy to identify the M-theory lifts
of all orientifold 2-planes we found above.  The O2$^-$ plane has RR
charge $-1/8$ and therefore lifts to an $(\BR^7 \times S^1)/\BZ_2$
orbifold where both $p_+$ and $p_-$ are of type A. The $\tO$ lifts to
an $(\BR^7 \times S^1)/\BZ_2$ orbifold where both $p_+$ and $p_-$ are
of type B.  Finally, an O2$^+$ orientifold has RR charge $1/8$ and
therefore must lift to an orbifold where one of the points $p_+,p_-$
is of type A and another is of type B.  

Let us now justify the claim that both O2$^-$ and $\tO$ planes have
trivial $H^3(\RP6,\tZ)$ torsion, while O2$^+$ has nontrivial
torsion. To this end one has to compute the flux of the NS 2-form $B$
through a homology 2-cycle given by $x_3=x_4=x_5=x_6=0$. This is the
same as computing the flux of the M-theory 3-form $C$ through the
M-theory 3-cycle given by the same formulas. This 3-cycle is
homologous to the sum of two small 3-cycles at the points $p_+$ and
$p_-$. The fluxes through these 3-cycles for points of type A and B
are $0$ and $\pi$, respectively.  Recalling the M-theory lifts of the
O2 planes, we get the desired result.

\subsec{Duality of $\CN=8$ gauge theories}

Let us put together the results of the previous two subsections and
obtain dualities between $\CN=8$ theories with gauge groups
$O(2N),O(2N+1)$, and $Sp(2N)$.  These field theories have special
points $m_+,m_-$ on their moduli space where the metric has 
orbifold singularities. At these special points the theories flow to an
SCFT described by $N$ membranes near an $\BR^8/\BZ_2$ orbifold.  The
M-theory orbifold can be of type A or B, and so we can get two distinct 
SCFT's which will call A and B-models. We showed above that 

(i) The $O(2N)$ theory flows to the A-model at both $m_+$ and $m_-$.

(ii) The $O(2N+1)$ theory flows to the B-model at both $m_+$ and $m_-$.

(iii) The $Sp(2N)$ theory flows to the A-model at one point and to the
B-model at another point. Which point is which is a matter of
convention in field theory. In string theory this information is
presumably encoded in the cohomology class of the RR 4-form field
strength.

Thus the $Sp(2N)$ theory is dual to both the $O(2N)$ theory and the 
$O(2N+1)$
theory in the vicinity of certain singular points in the moduli space. 
The IR equivalence of the 
$Sp(2N)$ and $O(2N+1)$ at the origin of the moduli space has been 
previously shown in \rsethi.

An interesting special case arises if we set $N=1$. The $O(2)$ theory
is a $\BZ_2$ orbifold of the $SO(2)$ theory which is free. Thus for
$N=1$ the A-model is a an orbifold of a free SCFT. The $O(3)$ theory,
on the other hand, flows to an interacting theory at both $m_+$ and
$m_-$ \rseiberg, hence the B-model is interacting. Finally, with a
suitable choice of the cohomology class of the RR 4-form field
strength the $Sp(2)=SU(2)$ theory flows to an interacting fixed point
at $m_+$ and to a free orbifold at $m_-$. This agrees with the results
of \rseiberg.

\newsec{${\cal N}=4$ supersymmetric gauge theories in three dimensions}

In this section we will explore IR dualities for $d=3,\ \CN=4$
models. Even though one expects a myriad of $\CN=4$ dual pairs, we
will restrict ourselves to the simplest singularities in IIA and
M-theory. We will be interested in the following two theories:

\item{(1)} $U(2N)$ with two hypermultiplet in the antisymmetric tensor
representation and two hypermultiplets in the fundamental. For reasons
explained below, we shall refer to this theory as the $(A,B)$-model.
\item{(2)} $Sp(2N)\times Sp(2N)$ with a hypermultiplet in the
$({\bf 2N,2N})$, a hypermultiplet in the $({\bf 2N,1})$ and a 
hypermultiplet in the $({\bf 1,2N})$. We shall refer to this theory as 
the $(B,B)$-model.

\subsec{IIA brane configurations}

The IIA backgrounds which the D2-branes probe are given by
free worldsheet CFT's. The common ingredients in the CFT's are:

\medskip

\item{(1)}
An orientifold 6-plane which is a point in $x^{7,8,9}$. The
corresponding orientifold action, $\Omega'$, will be specified more
precisely below.

\item{(2)}
A pair (on the base space) of D6-branes branes parallel 
to the $O6$-plane.

\item{(3)}
A $\BZ_2$ orbifold action on the coordinates 
$x^{3,4,5,6}$ (denoted by $R$).

\medskip

In all cases we will take the orientifold plane to 
be the $O6^-$ plane which has D6-brane charge\foot{Charges will
always be counted on the base space.} $-2$. This implies that we can take
\pg
\eqn\gomgs{\gamma_{\Omega',6}=1_{4\times 4},\ \ \ 
 	   \gamma_{\Omega',2}=
	\biggl(\matrix{0&1_{2N\times 2N}\cr -1_{2N\times 2N}&0}\biggr),}
where there are $N$ physical D2-branes.
 
We now have the freedom to choose \joet\ the sign $\eta$ in the equation
\eqn\gsgn{\gamma_{\Omega'R,6}=\eta \gamma_{\Omega'R,6}^T,\ \ \ 
	  \gamma_{\Omega'R,2}=-\eta\gamma_{\Omega'R,2}^T.}  
This choice is correlated with the action of $\Omega'$ on the twisted
sector states of $R$. The twisted sector states are a RR vector field
$A^{(1)}_{0,1,2}$, 3 NSNS scalars $\phi_i$, a RR vector $A^{(2)}_{7,8,9}$
and an NSNS scalar $\phi_0$. For $\eta=-1$ $1+\Omega'$ projects (at the
intersection of the orientifold and orbifold) onto $A^{(1)}$ and
$\phi_i$, and for $\eta=1$ it projects onto $A^{(2)}$ and $\phi_0$.

The orbifold with $\eta=-1$ is T-dual to the Gimon-Polchinski model. A
convenient choice for the projection matrices is:
\eqn\gmab{
\gamma_{R,6}=
{\biggl(\matrix {&1_{2\times 2}&\cr-1_{2\times 2}&}\biggr)},\ \ 
\gamma_{R,2}=
{\biggl(\matrix {&\sigma_1\cr-\sigma_1&}\biggr) }} 
This gives rise to the $(A,B)$ model on the D2 worldvolume.  

For $\eta=1$, a convenient choice of matrices is:  
\eqn\gmbb{
\gamma_{R,6}=
{\biggl(\matrix{1_{2\times 2}&\cr &-1_{2\times 2}}\biggr)},\ \ 
\gamma_{R,2}=
{\biggl(\matrix{&i\sigma_2\cr-i\sigma_2&}\biggr) }} 

This gives rise to the $(B,B)$ model on the D2 worldvolume. 
$\gamma_{R,2}$ has to be traceless in order to cancel an
unphysical twisted tadpole.

Before the $R$ projection, the choice \gomgs\ determines the Lie
algebra of the gauge symmetry on the D6-branes to be $so(4)$. Choosing
$\gamma_{R,6}$ is equivalent to specifying a monodromy of a flat
connection on $\BR^4/\BZ_2$ with the origin removed. It is easy to see
that in the $(A,B)$ model the gauge bundle does not admit a vector
structure \blsspw (see also \ken), while in the $(B,B)$ model it
admits a vector structure but no spinor structure. This suggests that
the bundle is actually an $SO(3)\times SO(3)$ bundle.

\subsec{M-theory interpretation}

Both the $(A,B)$ and $(B,B)$ models lift to N M2-branes probing
M-theory on $\BR^4/\BZ_2\times (\BR^3\times S^1)/\BZ_2$.  This
background has two points that are fixed by the entire orbifold group:
one at $x^{3..9}=0,\ \sigma=0$ and another at $x^{3..9}=0,\
\sigma=\pi$. As before, we will denote these points by $p_+$ and $p_-$.
The overall picture is that of two parallel 7-planes of $A_1$
singularities which are intersected by an orthogonal 7-plane of $A_1$
singularities. The local behavior near each intersection is therefore
of the form $\BR^4/\BZ_2\times \BR^4/\BZ_2$.  Since we have three
$A_1$ planes, one would expect to have three $su(2)$ gauge multiplets. A
pair of $su(2)$'s that lives on the parallel $A_1$'s corresponds to the
$so(4)$ multiplet on the D6-branes. The third $su(2)$, however, must
be broken by a Wilson line if we are to obtain a perturbative type IIA
model. This will explained in greater detail below.
 
The two parallel $A_1$ singularities wrap
$\BR^4/\BZ_2$. Supersymmetric vacua correspond to self-dual
$SO(3)\times SO(3)$ connections on $\BR^4/\BZ_2$. Instantons on this
space were discussed in \blsspw, and we remind the reader the salient
features of that discussion below. The remaining $A_1$ wraps
$(\BR^3\times S^1)/\BZ_2$ necessitating the analysis of $SO(3)$
bundles on this space. The situation is further complicated by the
fact that the monodromies on the the two spaces are correlated, a fact
which we explain below.

The net result will be that there are two types of $\BR^4/\BZ_2\times
\BR^4/\BZ_2$ singularities, which we will call $A$ and 
$B$-singularities. An astute reader no doubt anticipates that the $(A,B)$
model contains one of each singularities, while the
$(B,B)$ model contains two $B$-singularities. The astute reader is
correct.

\vskip0.15in
\noindent{\it 3.2.1. $SO(3)$ bundles on $\BR^4/\BZ_2$}
\vskip0.05in

For our purposes it is sufficient to consider point-like instantons stuck 
at the fixed point of the orbifold. Arbitrary instantons can be 
obtained by combining such an object with ordinary instantons which are 
free to roam on $\BR^4/\BZ_2$. From the string theory point of view, 
ordinary instantons are simply D2-branes stuck to D6-branes; we can
always shrink them to zero size and move off the $A_1$ singularity.

The stuck instanton is flat everywhere except the origin.  Let us
denote $\BR^4/\BZ_2$ with the origin removed by $\tilde{\BR^4/\BZ_2}$.
The fundamental group of $\tilde{\BR^4/\BZ_2}$ is $\BZ_2$; flat
connections are classified by the conjugacy class of monodromy around
the generator of this $\BZ_2$. Up to conjugacy, there are two possible
monodromies, the trivial one and the one given by
\eqn\monU{U=\Biggl(\matrix{-1& & \cr &-1& \cr & & 1}\Biggr).}
If the monodromy is conjugate to $U$, the parallel transport for spinors
of $SO(3)$ cannot be defined: this is an $SO(3)$ connection without
spinor structure.

An alternative way of thinking about the stuck instanton is to blow-up
the singularity slightly, making $\BR^4/\BZ_2$ into a Eguchi-Hanson
space $M_{EH}$. Then the instanton becomes a bona fide self-dual
connection on $M_{EH}$. In the case of trivial monodromy it is a
trivial connection.  In the case of nontrivial monodromy it is a
connection with a nontrivial second Stiefel-Whitney class. Such a
connection can be constructed by embedding a particular $U(1)$
instanton on $M_{EH}$ into $SO(3)$ \blsspw.  Its topological charge is
$1/8$ of the charge of a ``free'' $SO(3)$ instanton.

\vskip0.15in
\noindent{\it 3.2.2. M-theory on $\BR^4/\BZ_2\times \BR^4/\BZ_2$}
\vskip0.05in

As explained above, the local geometry near either $p_+$ or $p_-$ is that
of two orthogonal planes of $A_1$ singularities, i.e. 
$\BR^4/\BZ_2\times \BR^4/\BZ_2$. We will label the coordinates of the
first $\BR^4/\BZ_2$ by $x^{\alpha,\beta,..}$ and of the second one by
$y^{\mu,\nu,..}$. It follows that we have two $SO(3)$ bundles, one living
at $x=0$ and another one at $y=0$. Naively the bundles appear independent,
but we as we will now see this is not the case.

{}From the previous subsection we know that after a slight blow-up a
nontrivial $SO(3)$ bundle is characterized by a self-dual field
strength $F$ in some $U(1)$ subgroup of $SO(3)$. This $U(1)$ subgroup
can be chosen so that the $U(1)$ gauge field $A(y)$ corresponds to a
$C$-field of the form $C(x,y)\sim A(y)\wedge w(x)$, where $w(x)$ is
the self-dual 2-form on $M_{EH}$ representing an integer cohomology
class. Thus a non-trivial $F$ corresponds to turning on a field
strength $G=dC$ of the form
\eqn\gflux {{G\over 2\pi}= w(x)\wedge w(y),} 
i.e. to a flux of $G$ through the 4-cycle dual to $w(x)\wedge w(y)$.
The constant of proportionality in  \gflux\ has been fixed by requiring
that $G/(2\pi)$ represent an integer cohomology class.

Note that this condition is symmetric under the interchange of the
two $\BR^4/\BZ_2$'s. This implies that if we have a nontrivial
$SO(3)$ monodromy on one $\BR^4/\BZ_2$, then there is also a
similar nontrivial monodromy on the other one. Therefore there are only
two types of $\BR^4/\BZ_2\times \BR^4/\BZ_2$ orbifolds in M-theory.
We will call the one with trivial monodromies an $A$-singularity, and
the one with both monodromies equal to $U$ a $B$-singularity.

For future use, let us compute the membrane charge of both types of
$\BR^4/\BZ_2\times \BR^4/\BZ_2$ orbifolds.  For the $A$-singularity
the charge comes from a $C\wedge X_8(R)$ interaction in $d=11$
supergravity, while for the $B$-singularity there is also a
contribution from the Chern-Simons term $C\wedge G\wedge G$. The
gravitational contribution to the charge is simply $-\chi/24$, where
$\chi$ is the integral of the Euler density. To compute $\chi$ we
consider a compactified orbifold $T^4/\BZ_2\times T^4/\BZ_2$ whose
Euler number is $2^8$ times the Euler number of $\BR^4/\BZ_2\times
\BR^4/\BZ_2$. The compactified orbifold can be blown up to $K3\times
K3$, so its Euler number is $24^2$.  It follows that the integral of
the Euler density for $\BR^4/\BZ_2\times
\BR^4/\BZ_2$ is $9/4$, and the gravitational contribution to the 
membrane charge 
is $-3/32$. When the $G$ field of the form \gflux\ is switched on,
there is an additional contribution to the charge
$${1\over 2} \int {G\over 2\pi}\wedge{G\over 2\pi}.$$
Since 
$$\int_{M_{EH}}w(x)\wedge w(x)=-{1\over 2},$$
this contribution is equal to $1/8$. Thus the membrane charges of
$A$ and $B$-singularities are $-3/32$ and $1/32$, respectively.

\vskip0.15in
\noindent{\it 3.2.3. $SO(3)$ bundles on $(\BR^3\times S^1)/\BZ_2$}
\vskip0.05in

Our orbifold $(\BR^3\times S^1)/\BZ_2\times \BR^4/\BZ_2$ contains a
submanifold of $A_1$ singularities which wraps $(\BR^3\times
S^1)/\BZ_2$. Thus we need to understand self-dual $SO(3)$ bundles on
this space. When the orbifold singularities in this space are removed,
the question reduces to the analysis of flat $SO(3)$ bundles. These
are classified by homomorphisms from the fundamental group to $SO(3)$
modulo conjugation.

\fig{Calculation of the fundamental group of 
$(\BR^3\times S^1)/\BZ_2$ with two singular points
deleted.}{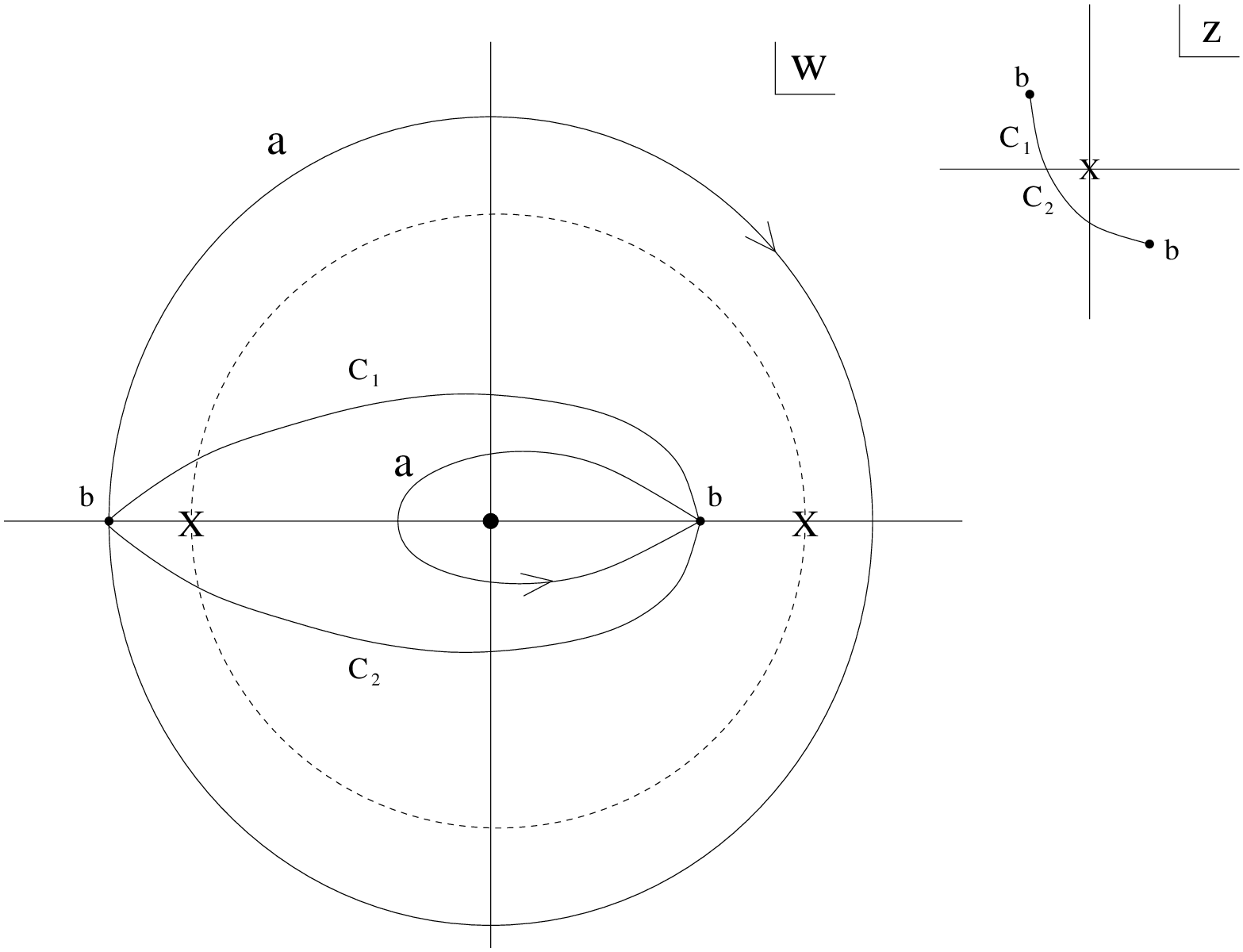}{3.7in}

The computation of the fundamental group is illustrated in Figure 1. In
this figure $z=x^1+ix^2$ and $w=e^{x^3+ix^4}$, with $x^{1..3}$ being the
affine coordinates on $\BR^3$, and $x^4$ being the angular coordinate
on $S^1$ with period $2\pi$. The origin of the
$w$-plane is excised for every value of $z$. The $\BZ_2$ projection acts
as $z\rightarrow -z,\ w\rightarrow 1/w$. The points $(w,z)=(1,0)$ and
$(-1,0)$, denoted by ``x'', are the fixed points of this action and
are removed as well.
It is clear from the figure that the generators $c_1,c_2,a$ of the
fundamental group satisfy
\eqn\fndgrp{c_1c_2=a,\ c_1^2=c_2^2=1,\ c_1ac_1=a^{-1}.} 
$c_2$ can be eliminated using these relations. However, it is useful
to keep it since Wilson lines along $c_1$ and $c_2$ measure the
monodromy around the two (symmetric) fixed points.  The homomorphisms
from this fundamental group to $SO(3)$ fall into three classes:
\item{(1)}
$${W(c_1)=W(c_2)=W(c_3)=1_{3\times 3},}$$
\item{(2)} 
$${W(c_1)=1_{3\times3},\ W(c_2)=U,\ W(a)=U,}$$
\item{(3)}
$${W(c_1)=U,\ W(c_2)=\CR(\phi)U\CR(\phi)^{-1},\ W(a)=U\CR(\phi)U
\CR(\phi)^{-1},}$$

\noindent where 
$$ \CR(\phi)=\Biggl(\matrix{1& & 
\cr &\cos{\phi\over 2}&\sin{\phi\over 2}\cr & -\sin{\phi\over 2}
 &\cos{\phi\over 2}}\Biggr).$$
\noindent Note that the third class is a one-parameter family indexed 
by $\phi\in
[0,\pi]$.

\vskip0.15in
\noindent{\it 3.2.4. M-theory on 
$(\BR^3\times S^1)/\BZ_2\times  \BR^4/\BZ_2 $}
\vskip0.05in

Putting all of this together, we come to the following classification of
$(\BR^3\times S^1)/\BZ_2\times  \BR^4/\BZ_2 $ orbifolds in M-theory.
There are three types of such orbifolds corresponding to three types
of flat $SO(3)$ bundle on $(\BR^3\times S^1)/\BZ_2$. Picking a particular
type of the bundle also fixes the monodromy of the
flat $SO(3)\times SO(3)$ bundle which lives on $\BR^4/\BZ_2$. For the
three choices described in the previous subsection the $SO(3)\times SO(3)$
monodromies are
\item{(1)} $(1_{3\times3},1_{3\times3})$
\item{(2)} $(1_{3\times3},U)$
\item{(3)} $(U,U)$

In the first case there are two $A$-singularities, in the second case
there is one $A$ and one $B$-singularity, and in the third case there
are two $B$-singularities.  In the first case $SO(3)\times SO(3)$
gauge group is not broken by the monodromy, in the second case it is
broken down to $U(1)\times SO(3)$, and in the third case it is broken
down to $U(1)\times U(1)$.  It is also important to know how the
monodromies on $(\BR^3\times S^1)/\BZ_2$ break the $SO(3)$ which lives
there. The unbroken group consists of all elements of $SO(3)$ which
commute with all the monodromies. In the $(A,A)$ case the $SO(3)$ is
unbroken, and in the $(A,B)$ it is broken down to $U(1)$. In the
$(B,B)$ case the $SO(3)$ is completely broken for generic values of
the parameter $\phi$, while for $\phi=0,\pi$ there is a residual
$U(1)$.

\vskip0.15in
\noindent{\it 3.2.5. Relation to IIA orientifolds}
\vskip0.05in

In the weak coupling limit the two parallel planes of $A_1$
singularities in M-theory become an $O6^-$ plane and a pair of
D6-branes. The $SO(3)\times SO(3)$ gauge bundle becomes the gauge
bundle on D6-branes.  As discussed in subsection 3.1, for the $(A,B)$
and $(B,B)$ orientifolds the gauge bosons on D6-branes are in the
adjoint of $u(2)$ and $u(1)\times u(1)$, respectively. Hence the
$(A,B)$ orientifold must be the weak coupling limit of the $(A,B)$
singularity in M-theory, while the $(B,B)$ orientifold comes from the
$(B,B)$ singularity.

Note that there is no IIA orientifold corresponding to the $(A,A)$
singularity. The reason is that in the $(A,A)$ case there is an
unbroken nonabelian gauge group living at the origin of the $\BR^4/\BZ_2$
orbifold. This gauge group is nonperturbative from the IIA point of view,
hence the corrresponding type IIA
background cannot be described by a free worldsheet CFT. As explained in
\aspinwall, perturbative orbifolds avoid gauge symmetry enhancement
by assigning a nonzero expectation value to a certain scalar in the
twisted NSNS sector. In the M-theory language, the breaking of the
symmetry is due to an $SO(3)$ Wilson line along the M-theory circle;
the above-mentioned scalar parametrizes its eigenvalues. In the
$(A,A)$ case the Wilson line $W(a)$ is frozen at $1$, so no
description based on a free orbifold is possible. In the $(A,B)$ case
the Wilson line is frozen at a nontrivial value which breaks $SO(3)$
down to $U(1)$. In the $(B,B)$ case the Wilson line is parametrized by
a real variable $\phi\in [0,\pi]$ which is identified with the twisted
NSNS scalar $\phi_0$ in IIA (see subsection 3.1). The perturbative IIA
construction picks a particular value for the VEV of $\phi_0$; one can
argue that this value is $\pi/2$.

As a check of this identification of M-theory orbifolds and IIA
orientifolds, let us compare their membrane charges. According to
subsection 3.2.2 the M2-brane charges of $(A,B)$ and $(B,B)$ orbifolds
are $-1/16$ and $1/16$, respectively.  The membrane charge of
perturbative IIA orientifolds can be determined by studying
tadpoles. Alternatively, we can make use of the fact that the $(A,B)$
model is T-dual to the Gimon-Polchinski model \pg, while the $(B,B)$
model is T-dual to the orientifold constructed in \joet. The charge of
the former is $-1/2$, while that of the latter is $1/2$. T-duality
along three directions parallel to the orientifold planes reduces the
charges by a factor of $8$, giving $-1/16$ and $1/16$.  Thus we find
complete agreement between the perturbative IIA computation and the
supergravity computation in $d=11$.

\subsec{Duality of $\CN=4$ gauge theories}

From the M-theory description it follows that the moduli space of both
$(A,B)$ and $(B,B)$ gauge theories is an orbifold ${\rm
Sym}((\BR^3\times S^1)/\BZ_2\times \BR^4/\BZ_2 )^N.$ This orbifold has
$2^N$ fixed points.  We denote by $m_+,m_-$ the two points on the
moduli space corresponding to all M2-branes at $p_+$ or $p_-$. Just as
in the $\CN=8$ case, it is sufficient to examine the theory at
$m_+,m_-$.

The discussion in subsection 3.2 implies that the $(B,B)$ model flows
to the same IR fixed point both at $m_+$ and $m_-$. This fixed point
has a conserved $u(1)$ current. The $(A,B)$ model flows to
inequivalent fixed points at $m_+$ and $m_-$. One of them has a
conserved $u(1)$, while the other one has an $su(2)$ current
algebra. The fixed point with the $u(1)$ current is the same as the
fixed point to which the $(B,B)$ model flows.

Classically, the full gauge symmetry is restored when the D2-branes
sit on top of the orientifold plane, irrespective of the VEVs of the
dual photons.  Therefore the IR equivalence of $(A,B)$ and $(B,B)$
models in the vicinity of orbifold singularities is a
quantum-mechanical phenomenon.

There exists yet another $\CN=4$ gauge theory which flows to the fixed
point with a conserved $u(1)$ and therefore is IR dual to both $(A,B)$
and $(B,B)$ models. It is a $U(N)\times U(N)$ theory with a four
hypermultiplets in the representations $({\bf N},{\bf \bar{N}}), ({\bf
\bar{N}},{\bf N}), (\Bone,{\bf N})$, and $({\bf N},\Bone)$. This
theory arises on D2-branes probing two D6-branes wrapped on
$\BR^4/\BZ_2$. If $\BZ_2$ acts on D2 Chan-Paton labels by
\eqn\gmorb{
\gamma_{R,2}=
{\biggl(\matrix{1_{N\times N}&\cr & -1_{N\times N}}\biggr)},}
cancellation of unphysical tadpoles requires
\eqn\gmorbb{
\gamma_{R,6}=
{\biggl(\matrix {1&\cr & -1}\biggr)}.}  It is easy to see that the
theory on the probes has the gauge group and matter content described
above. This IIA background lifts to $M_{TN2}\times \BR^4/\BZ_2$ in
M-theory, where $M_{TN2}$ is a two-center Taub-NUT space with
coincident centers. $M_{TN2}$ is topologically equivalent to
$\BR^4/\BZ_2$, therefore we are dealing with an $\BR^4/\BZ_2\times
\BR^4/\BZ_2$ orbifold of M-theory. Furthermore, from \gmorbb\ we see
that the $\BZ_2$ monodromy of the D6 bundle breaks the D6 gauge group
from $U(2)$ down to $U(1)\times U(1)$. From the M-theory point of
view, the diagonal $U(1)$ comes from the untwisted sector, while the
difference of the two $U(1)$'s comes from the twisted sector.  Recall
now that for $A$ and $B$ singularities the twisted sector gauge group
is $SO(3)$ and $U(1)$, respectively. We conclude that the IIA orbifold
with Chan-Paton matrices as in \gmorb,\gmorbb\ lifts to a singularity
of type $B$. It follows that at the origin of the moduli space the
probe theory is IR dual to both $(B,B)$ and $(A,B)$ models.

\bigbreak\bigskip\bigskip\centerline{{\bf Acknowledgements}}\nobreak

We would like to thank J.~Park, N.~Seiberg, S.~Sethi, and E.~Witten for 
helpful discussions. We are especially grateful to K.~Intriligator for
patient explanations of his work.
The work of M.B. is supported by NSF grant
PHY-9513835; that of A.K. by DOE grant DE-FG02-90ER40542.

\listrefs

\end